\documentclass[11pt, letterpaper]{article}
\usepackage[T1]{fontenc}
\usepackage{amsmath,amssymb}  
\usepackage{amsthm}
\usepackage{bm}      
\usepackage{dsfont}  

\usepackage[american]{babel}

\theoremstyle{remark}
\newtheorem*{rmk}{Remark}

\DeclareMathOperator{\diag}{diag}
\DeclareMathOperator{\rank}{rank}

\usepackage{geometry}
\usepackage{graphicx}
\usepackage{float}
\usepackage{caption}
\usepackage{paralist}    
\usepackage{footmisc}  
\usepackage{dcolumn}  

\usepackage[parfill]{parskip}    
\usepackage[hyperfootnotes=false,%
   colorlinks=true,citecolor=blue,urlcolor=blue,linkcolor=red%
   ]{hyperref}

\begin{document}

\tolerance=550  
\emergencystretch=2ex 

\thispagestyle{empty}
\begin{center}

\vspace*{0.5in}
\parbox{4.5in}{\LARGE
\centering
Notional portfolios and normalized linear returns
}

\vspace*{0.5in}
Vic Norton\\
Department of Mathematics and Statistics\\
Bowling Green State University\\
\url{mailto:vic@norton.name}\\
\url{http://vic.norton.name}

\vspace*{0.25in}
28-Apr-2011

\vspace*{1.0in}
ABSTRACT\\[2.5ex]
\begin{minipage}{4in}
The vector of periodic, compound returns of a typical investment
portfolio is almost never a convex combination of the return vectors of
the securities in the portfolio. As a result the ex post version of
Harry Markowitz's ``standard mean-variance portfolio selection model''
does not apply to compound return data. We propose using notional
portfolios and normalized linear returns to remedy this problem.
\end{minipage}
\end{center}

\setcounter{page}{0}


\clearpage
\section{The ex post standard model}
\label{stdmodel}

Let us paraphrase the description of the ``standard mean-variance
portfolio selection model'' (\cite[pp. 3--5]{Markowitz:1987wd}) for
ex post return data.

Given an $m \times n$ matrix $R = [\mathbf{r}_1,\ldots,\mathbf{r}_n]$ of
successive periodic returns ($m$ returns for each of $n$ securities), an
investor is to choose the proportions $\mathbf{p} = [p_1,\ldots,p_n]^T$
invested in each security, the proportions being subject to the
constraints $p_j\ge0~ (j=1,\ldots,n),~ \sum_{j=1}^n p_j = 1.$ We assume
that the periodic returns, $\mathbf{r}_P\in\mathds{R}^m$, of the
corresponding \emph{investment portfolio} satisfy the \emph{linear
hypothesis}
\begin{equation}\label{linear_hypothesis}
  \mathbf{r}_P = \sum_{j=1}^n \mathbf{r}_j p_j = R \mathbf{p}.
\end{equation}
We also assume that \emph{mean} or \emph{expected} return is a linear
function of periodic return: that the expected return of security $k$ is
given by $e_j=\bm{\omega}^T\bm{r}_j$ for $j=1,\dots,n$. Here the weight
vector $\bm{\omega}\in\mathds{R}^m$ should satisfy $\omega_i > 0~
(i=1,\dots,m)$ and $\sum_{i=1}^m\omega_i = 1.$ Typically $\omega_i =
1/m$ for $i=1,\dots,m$:~ every periodic return of a given security
contributes equally to its expected return.

Under these assumptions, the expected return of
the investment portfolio is
\begin{equation}\label{expected_return}
  e_P = \sum_{j=1}^n e_j p_j = E \mathbf{p},
\end{equation}
with~ $E = [e_1,\ldots,e_n] = \bm{\omega}^T R$, and the \emph{variance}
of portfolio return is
\begin{equation}\label{variance}
  v_P = \sum_{j=1}^n\sum_{k=1}^n v_{jk} p_j p_k = \mathbf{p}^T V \mathbf{p},
\end{equation}
where the $n\times n$ \emph{covariance} matrix ~$V = [v_{jk}]$~ is given
by
\begin{equation}\label{covariance}
  v_{jk}= \sum_{i=1}^m \omega_i z_{ij}z_{ik}
  \quad (j, k = 1,\ldots n),
\end{equation}
the deviation vectors~ $\mathbf{z}_j\in\mathds{R}^m~ (j=1,\ldots,n)$~
being defined by
\begin{equation}\label{deviation_vectors}
  \mathbf{z}_j = \mathbf{r}_j - \mathbf{1}_m e_j,
\end{equation}
with $\mathbf{1}_m\in\mathds{R}^m$ representing the constant return
vector of all 1's.


\section{An investment portfolio}
\label{investmentportf}

We will illustrate the ideas in this paper with computations based on
data for five iShares exchange traded funds (ETFs) over the year
2010. The funds are

\hspace*{4ex}
\begin{minipage}{5in}\label{fivefunds}
\begin{compactenum}
  \item \texttt{IEF} -- iShares Barclays 7-10 Year Treasury Bond Fund
  \item \texttt{IWB} -- iShares Russell 1000 Index Fund
  \item \texttt{IWM} -- iShares Russell 2000 Index Fund
  \item \texttt{EFA} -- iShares MSCI EAFE Index Fund
  \item \texttt{EEM} -- iShares MSCI Emerging Markets Index Fund
\end{compactenum}
\end{minipage}

Table \ref{portf} shows the 2010 performance of a hypothetical
investment portfolio in the five funds. The portfolio, \texttt{PORTF},
is static in the sense that no sales or additional purchases were made
during the one year period. The increases in the shares of its component
funds are entirely due to (automatic) dividend reinvestment.

\begin{table}[th]
  \centering
  \caption[portf]{\label{portf}%
    A (static) investment portfolio
    }
  { \fontsize{10}{13}\selectfont
    \newcolumntype{e}{D{.}{.}{2}}
    \newcolumntype{f}{D{.}{.}{3}}
    \newcolumntype{g}{D{,}{,}{3}}
  \begin{tabular}{|c|eeg|f|}
    \hline\rule[-0.7ex]{0ex}{2.9ex}
    fund & \multicolumn{1}{c}{shares} & \multicolumn{1}{c}{price}
      & \multicolumn{1}{c}{value} & \multicolumn{1}{|c|}{proportion} \\
    \hline
    \multicolumn{5}{|c|}{\rule[0ex]{0ex}{2.0ex}
    At the close of Thursday, 2009-12-31} \\
    \hline\rule[0ex]{0ex}{2.0ex}
      \texttt{IEF} & 395.03 & 88.60 & 35,000 & 35.00\% \\
      \texttt{IWB} & 652.42 & 61.31 & 40,000 & 40.00\% \\
      \texttt{IWM} & 0.00 & 62.44 & \multicolumn{1}{c|}{0} & 0.00\% \\
      \texttt{EFA} & 452.24 & 55.28 & 25,000 & 25.00\% \\
      \texttt{EEM} & 0.00 & 41.50 & \multicolumn{1}{c|}{0} & 0.00\% \\
    \hline\rule[0ex]{0ex}{2.0ex}
      \texttt{PORTF} & & & 100,000 & 100.00\% \\
    \hline
    \multicolumn{5}{|c|}{\rule[0ex]{0ex}{2.0ex}
    At the close of Friday, 2010-12-31} \\
    \hline\rule[0ex]{0ex}{2.0ex}
      \texttt{IEF} & 407.97 & 93.82 & 38,276 & 34.25\% \\
      \texttt{IWB} & 664.51 & 69.86 & 46,422 & 41.54\% \\
      \texttt{IWM} & 0.00 & 78.24 & \multicolumn{1}{c|}{0} & 0.00\% \\
      \texttt{EFA} & 464.48 & 58.22 & 27,042 & 24.20\% \\
      \texttt{EEM} & 0.00 & 47.64 & \multicolumn{1}{c|}{0} & 0.00\% \\
    \hline\rule[0ex]{0ex}{2.0ex}
      \texttt{PORTF} & & & 111,741 & 100.00\% \\
    \hline
  \end{tabular}
  }
\end{table}

\medskip
\begin{rmk}
All results in this paper are based on data from the spreadsheet file
`\href{http://vic.norton.name/finance-math/notionportf/adjclose5_2010.csv}%
  {adjclose5\_2010.csv}'.
Computations were done in double precision arithmetic with results
rounded for presentation. Consequently certain numbers (e.g., the sums
of the 2010-12-31 values and proportions in Table \ref{portf}) may be
seem to be off by 1 in the last digit.
\end{rmk}

\bigskip
In the standard model of portfolio selection an investor is to ``choose
proportions invested in each security.'' But which proportions are
appropriate for the portfolio \texttt{PORTF} of Table \ref{portf}---the
2009-12-31-closing proportions, the 2010-12-31-closing proportions, or
some set of market-day-closing proportions in between?

To examine this question we need a definition of what we mean by an
``investment portfolio.'' Say you invest a certain amount of money in
$n$ funds, and you let it ride, with all dividends reinvested
automatically. At the close of market day one your investment is worth
so much. At the close of the market day two it has another, probably
different, value. And so it goes, market day after market day, value
after value. The growth of your investment portfolio is described by
this whole sequence of market-day-closing values indexed by the
successive market days of your investment. These market-day-closing
values are \emph{adjusted closing prices} for your portfolio.


\section{Adjusted closing prices and notional shares}
\label{adjcloseprices}

We refer to any vector of positive numbers $\mathbf{x}=[x_i]$, indexed
by successive market days $i$, as (a vector of) \emph{adjusted closing
prices} for a given security if the ratio $x_i/x_{i-1}$ measures the
growth in value of an investment in the security from the close of
market day $i-1$ to the close of day $i$, assuming dividends are
automatically reinvested. Ordinarily $x_i/x_{i-1} = c_i/c_{i-1}$, where
$\mathbf{c}=[c_i]$ is the vector of closing prices for the security.
However, on ex-dividend-days $i$,
\begin{equation}\label{adjclosedef}
  x_i/x_{i-1} =
  \begin{cases}
    c_i/(c_{i-1} - d) &\text{if the dividend is $d$ dollars per share;}\\
    (1+s)c_i/c_{i-1}  &\text{if the dividend is $s$ shares per share;}\\
    \tau c_i/c_{i-1} &\text{if shares are split $\tau:1$.}
  \end{cases}
\end{equation}

Clearly, if $\mathbf{x}$ is a vector of adjusted closing prices for a
security, then $\mathbf{y} = \mathbf{x}\lambda$ is a vector of adjusted
closing prices for the same security for any $\lambda>0$. Moreover, all
adjusted closing price vectors for the same security and covering the
same time period can be expressed in the form $\mathbf{y} =
\mathbf{x}\lambda,~ \lambda>0$.

The spreadsheet file
`\href{http://vic.norton.name/finance-math/notionportf/adjclose5_2010.csv}%
  {adjclose5\_2010.csv}'.
contains adjusted closing prices for our five sample ETFs over the 253
market days from Thursday, 2009-12-31, through Friday, 2010-12-31. The
adjusted prices of each security are normalized at 100 on 2009-12-31.
See \cite{Norton:2010fk} for a more extensive discussion of adjusted
closing prices.

\bigskip
Let $X=[\mathbf{x}_1,\ldots,\mathbf{x}_n]$ be a matrix of adjusted
closing prices for $n$ securities over a certain time period.  Assume
the columns of $X$ are linearly independent ($\rank(X)=n$). Then
adjusted closing prices $\mathbf{x}_P$ can be defined for any (static)
investment portfolio $P$ in the $n$ securities by
\begin{equation}\label{xP}
  \mathbf{x}_P = \sum_{j=1}^n\mathbf{x}_j s_j = X\mathbf{s} \quad
\text{with}\quad s_j\ge0~ (j=1,\dots,n) ~\text{and}~ \sum_{j=1}^n s_j>0.
\end{equation}
Moreover, due to the independence of the $\mathbf{x}_j$, the
\emph{notional shares} $\mathbf{s} = [s_j]$ are uniquely determined by
the adjusted closing prices $\mathbf{x}_P$.

For example, choose a market day $i_0$, let $x_{i_0,j}$ be the closing
price of security $j$ on that day, and let $s_j$ be the number of shares
of security $j$ held in the portfolio on that day ($j=1,\dots,n$). Then
$x_{i_0,P}=\sum_{j=1}^nx_{i_0,j} s_j$ is the value of the portfolio at
the close of day $i_0$. Now using \eqref{adjclosedef}, with $d=0$ on
non-ex-dividend-days $i$, the $x_{i_0,j}$ can be extended to adjusted
closing prices for all $n$ securities over the whole time period
(\cite{Norton:2010fk}). The resulting $x_{iP}=\sum_{j=1}^nx_{ij} s_j$
are then the market-day-closing values of the investment portfolio $P$.

This argument shows that a vector of closing values, $\mathbf{x}_P$,
of any investment portfolio $P$ can be realized by equation
\eqref{xP}---with the specific matrix of adjusted closing prices,
$X$, and notional shares, $\mathbf{s}$, described in the argument.

Now suppose that
\begin{equation}\label{xylambda}
  \mathbf{y}_P=\mathbf{x}_P\lambda_P,\quad
  Y=X\diag(\bm{\lambda}),\quad \bm{\lambda}=[\lambda_1,\ldots,\lambda_n]^T,
\end{equation}
with $\lambda_P>0$ and $\lambda_j>0$ for $j=1,\ldots,n$. Then
$\mathbf{y}_P$ is a vector of adjusted closing prices for the portfolio
$P$, $Y$ is a matrix of adjusted closing prices for the $n$ securities
in $P$, and all matrices of adjusted closing prices for $P$ and its $n$
component securities can be represented this way. Moreover
\begin{equation}\label{yP}
  \mathbf{y}_P=\sum_{j=1}^n \mathbf{y}_jt_j
  =Y\mathbf{t} \quad\text{with}\quad
  \mathbf{t}=\diag(\bm{\lambda})^{-1}\mathbf{s}\,\lambda_P
\end{equation}
as a consequence of
\begin{equation*}
\mathbf{y}_P=\mathbf{x}_P\lambda_P=X\mathbf{s}\,\lambda_P
=X\diag(\bm{\lambda})\diag(\bm{\lambda})^{-1}\mathbf{s}\,\lambda_P
=Y\mathbf{t}.
\end{equation*}
Finally, the columns of $Y$ are linearly independent if and only if the
columns of $X$ are linearly independent ($\rank(Y)=\rank(X)$). Thus
equation \eqref{yP} is effectively equivalent to equation \eqref{xP},
and equation \eqref{xP} is valid in the general circumstances
described.\\

\begin{rmk}
One can also think of (\ref{xP}--\ref{yP}) as \emph{change of
coordinate} equations, with a fixed investment portfolio $P$ being
represented by different vectors of notional shares in different
adjusted-closing-price systems.
\end{rmk}


\section{Market-day-closing portfolios}
\label{closingportf}

Suppose an investment portfolio $P$ is described by equation \eqref{xP}.
Multiplying both sides of the equation by a positive constant if
necessary, we may assume that $\mathbf{x}_P$ is the vector of
market-day-closing values of the portfolio. Since the value of the
portfolio at the close of day $i$ is the sum of the values of its
component securities,
\begin{align}
  x_{iP} = \sum_{j=1}^n x_{ij}s_j
    &= x_{iP} \sum_{j=1}^n \left(\frac{x_{ij}}{x_{iP}}\right)s_j
    = x_{iP} \sum_{j=1}^n p^c_{ij} \label{pc1}\\
\intertext{with}
  p^c_{ij} &= \left(\frac{x_{ij}}{x_{iP}}\right)s_j. \label{pc2}
\end{align}

Apparently $p^c_{ij}$ is the proportion of security $j$ in the portfolio
$P$ at the close of day $i$. In particular $p^c_{ij}\ge0~
(j=1,\ldots,n)$ and $\sum_{j=1}^n p^c_{ij}=1$.

If $\mathbf{y}_P, Y$, and $\mathbf{t}$ also represent $P$, as in
\eqref{xylambda} and \eqref{yP}, then
\begin{equation*}
  \left(\frac{y_{ij}}{y_{iP}}\right)t_j
    = \left(\frac{x_{ij}\lambda_j}{x_{iP}\lambda_P}\right)
    s_j\left(\frac{\lambda_P}{\lambda_j}\right)
    = \left(\frac{x_{ij}}{x_{iP}}\right)s_j
    = p^c_{ij}.
\end{equation*}

Thus the proportions $p^c_{ij}$ of $\eqref{pc1}$ and $\eqref{pc2}$ are
independent of the adjusted closing prices and notional shares used to
represent $P$. We refer to the matrix $P^c = [p^c_{ij}]$, indexed by
market days $i$ and securities $j$, as the matrix of
\emph{market-day-closing portfolios} for the investment portfolio $P$.

Table \ref{portf40} shows end-of-week market-day-closing portfolios for
the portfolio \texttt{PORTF} of Table \ref{portf} over the last forty
weeks of 2010. The weeks are indexed by Fridays, even though Good
Friday, 2010-04-02, and observed Christmas Friday, 2010-12-24, were
market holidays. All forty of these closing portfolios are different. In
fact, the market-day-closing portfolios of \texttt{PORTF} over the 253
market days from 2009-12-31 though 2010-12-31 are all distinct. This
again begs the question---what proportions
$\mathbf{p}=[p_1,\ldots,p_n]^T$ represent the investment portfolio
\texttt{PORTF} in the standard portfolio selection model?


\section{Normalized adjusted closing prices and notional portfolios}
\label{notionalportf}

Table \ref{prices40a} shows weekly adjusted closing prices for the
securities and the investment portfolio of Table \ref{portf} for the
last 40 weeks of 2010. The prices of the securities and the portfolio
have been normalized at \$100 per notional share on 2009-12-31. As a
consequence, the notional shares of \texttt{PORTF} with respect to these
adjusted prices are the same as its [2009-12-31]-closing proportions:
\begin{equation}\label{portfa}
   \texttt{PORTF} = 35.00\%\times\texttt{IEF} + 40.00\%\times\texttt{IWB}
   + 0\%\times\texttt{IWM} + 25.00\%\times\texttt{EFA} + 0\%\times\texttt{EEM}.
\end{equation}

\begin{figure}[H]
  \centering
  \caption{\label{adjclosea}%
    [2009-12-31]-normalized adjusted closing prices}
  \includegraphics{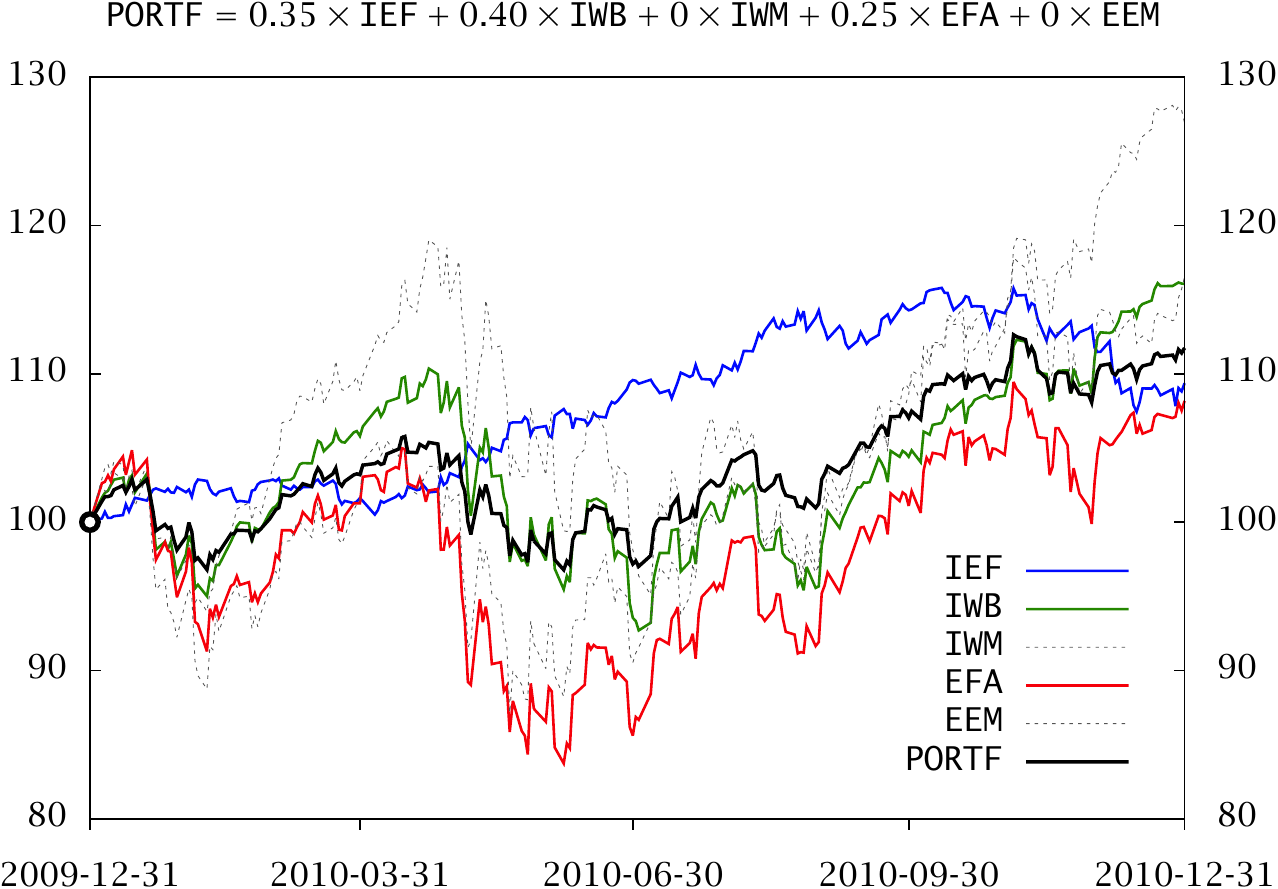}
\end{figure}

Figure \ref{adjclosea} shows the graphs of these adjusted closing prices
over the whole of 2010. The [2009-12-31]-normalization is apparent: all
graphs start at 100. We have not tried to distinguish \texttt{IWM} and
\texttt{EEM} in this figure since these securities are not components of
the investment portfolio \texttt{PORTF}.

Table \ref{prices40b} shows weekly adjusted closing prices for the same
securities and the same investment portfolio, but the adjusted prices in
Table \ref{prices40b} have been normalized in an entirely different way:
the average week-ending adjusted closing price over the last 13 weeks of
2010 is 100---for each security and the portfolio. The notional shares
of \texttt{PORTF} with respect to this system of adjusted closing prices
must also sum to one. In Table \ref{prices40b}
\begin{equation}\label{portfb}
   \texttt{PORTF} = 35.65\%\times\texttt{IEF} + 40.35\%\times\texttt{IWB}
   + 0\%\times\texttt{IWM} + 24.01\%\times\texttt{EFA} + 0\%\times\texttt{EEM}.
\end{equation}

\begin{figure}[H]
  \centering
  \caption{\label{adjcloseb}%
    $\alpha$-normalized adjusted closing prices}
  \includegraphics{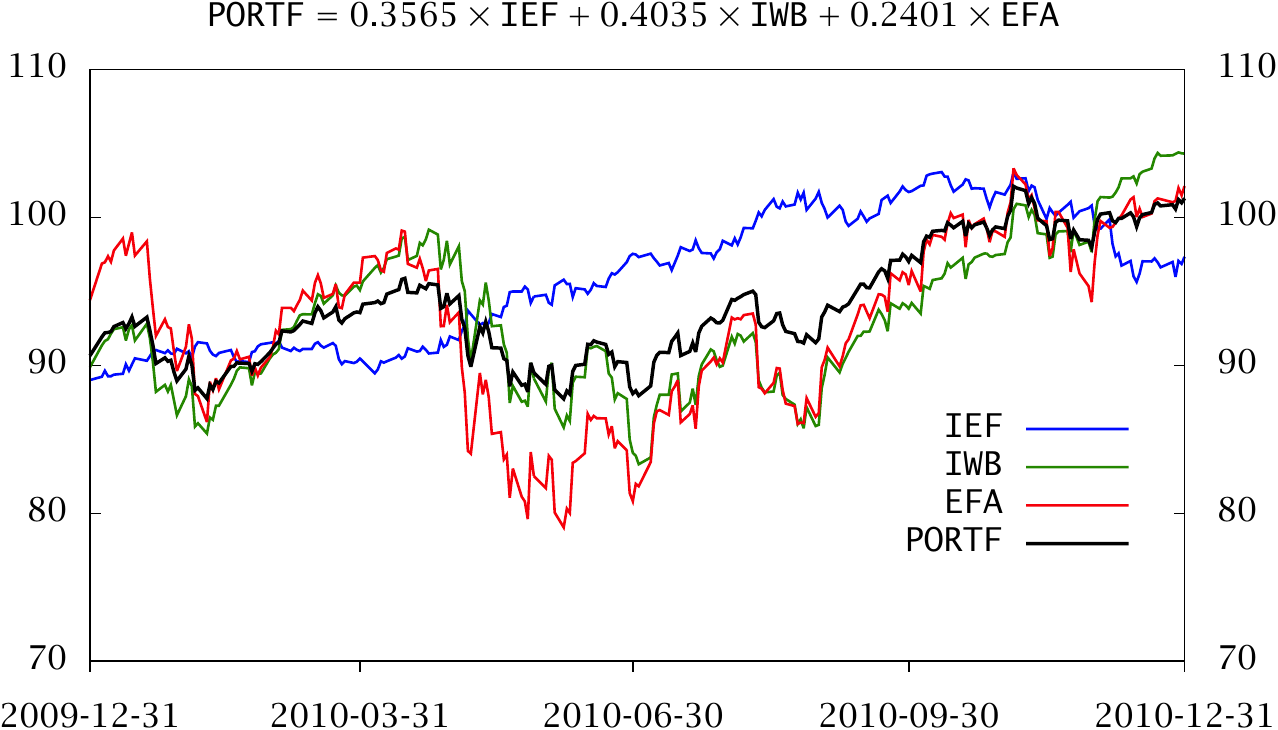}
\end{figure}

Figure \ref{adjcloseb} shows the graphs of these
last-13-week-normalized adjusted closing prices over 2010, but with
the graphs of \texttt{IWM} and \texttt{EEM} omitted. Each graph in
Figure \ref{adjcloseb} has exactly the same shape as the corresponding
graph in Figure \ref{adjclosea} in the sense that the two
adjusted-closing-price functions are positive multiples of one another.

We refer to the adjusted closing prices in Table \ref{prices40b} and
Figure \ref{adjcloseb} as being ``\mbox{$\alpha$-normalized},'' with
$\bm{\alpha}$ the market-day-averaging-vector defined by
\begin{equation}\label{alpha}
  \alpha_i =
  \begin{cases}
    \dfrac{1}{13} &\text{%
       \parbox{3in}{
         if $i$ corresponds to the last market day of
          one of the last 13 weeks of 2010;}
          }\\[1.5ex]
      ~\,0  &\text{otherwise.}
  \end{cases}
\end{equation}
The adjusted closing prices $[X,\mathbf{x}_P]$ of Table \ref{prices40b}
and Figure \ref{adjcloseb} are \emph{$\alpha$-normalized} at 100 in the
sense that $\bm{\alpha}^T[X,\mathbf{x}_P]=[100,\ldots,100]$.

\bigskip
We will say that a vector $\bm{\alpha}=[\alpha_i]$, indexed by
successive market days $i$, is a
\emph{market-day-averaging-vector}\label{marketdayavg} if
$\alpha_i\ge0$, for all $i$, and $\sum_i\alpha_i=1$. Given a
market-day-averaging-vector $\bm{\alpha}$, any vector of adjusted
closing prices $\mathbf{x}$ has a unique
"\emph{$\alpha$-normalized}-at-100" counterpart $\mathbf{x}^\alpha$---a
vector of adjusted closing prices for the same security that satisfies
$\bm{\alpha}^T\mathbf{x}^\alpha=100$. Clearly
\begin{equation}\label{xalpha}
  \mathbf{x}^\alpha = \mathbf{x}\cdot\dfrac{100}{\bm{\alpha}^T\mathbf{x}}.
\end{equation}
Just as the adjusted closing prices of Table \ref{prices40b} and Figure
\ref{adjcloseb} are $\alpha$-normalized (at 100) for the $\bm{\alpha}$
described by \eqref{alpha}, the adjusted closing prices of Table
\ref{prices40a} and Figure \ref{adjclosea} are $\alpha$-normalized (at
100) for the $\bm{\alpha}$ that is 1 if $i\sim$ 2009-12-31 and 0 at
every other market day $i$.

We have ``normalized'' adjusted closing prices at 100 in this paper
because 100 is convenient. For instance, if the adjusted closing price
of a security or portfolio starts at 100 (as in Figure \ref{adjclosea}),
then the change in adjusted price at some later time is exactly the
percentage gain or loss from the start. But the value 100 is not really
essential to our arguments. Any other positive value would work just as
well, and we will delete the normalizing qualifier ``at 100'' in what
follows.

Given $\alpha$-normalized adjusted closing prices
$X^\alpha=[\mathbf{x}^\alpha_1,\ldots,\mathbf{x}^\alpha_n]$ and
$\mathbf{x}^\alpha_P$, for $n$-securities and an investment portfolio
$P$ in these securities, and assuming that $\rank({X^\alpha})=n$, the
notional shares of $\mathbf{x}^\alpha_P$ with respect to $X^\alpha$ are
uniquely determined by $\bm{\alpha}$ and sum to one. We denote the
vector of these proportions by
$\mathbf{p}^\alpha=[p^\alpha_1,\ldots,p^\alpha_n]^T$ and refer to
$\mathbf{p}^\alpha$ as the \emph{$\alpha$-notional portfolio} of $P$.
Thus
\begin{equation}\label{palpha}
  \mathbf{x}^\alpha_P = \sum_{j=1}^n\mathbf{x}^\alpha_j p^\alpha_j
    = X^\alpha\mathbf{p}^\alpha \quad
    \text{with}\quad p^\alpha_j\ge0~ (j=1,\dots,n)
    ~\text{and}~ \sum_{j=1}^n p^\alpha_j=1.
\end{equation}
This notional portfolio may \eqref{portfa} or may not \eqref{portfb} be
the same as any of the market-day-closing portfolios in $P^c$.

If $\bm{\beta}$ is a second market-day-averaging-vector, then
\begin{equation}\label{pa2pb}
  p^\beta_j = \left(
    \frac{\bm{\beta}^T\mathbf{x}^\alpha_j}
    {\bm{\beta}^T\mathbf{x}^\alpha_P}\right)
    p^\alpha_j \quad (j=1,\ldots,n,P)
\end{equation}
as a consequence of
$
\mathbf{x}^\beta_j=
  \mathbf{x}^\alpha_j\cdot[100/(\bm{\beta}^T\mathbf{x}^\alpha_j)]
$
and \eqref{yP}.


\section{Compound returns}
\label{compoundreturns}

Given a vector $\mathbf{x}$ of periodic (e.g., weekly) adjusted closing
prices for a particular security or portfolio, the percentage change in
value, $r_i$, of the security or portfolio, over the period from $i-1$
to $i$, is given by
\begin{equation}\label{cmpd}
  \text{periodically compounded return:}\qquad
  r_i = \frac{x_i}{x_{i-1}} - 1.
\end{equation} On the other hand, if one thinks of the adjusted closing
prices as growing according to the exponential model
$x_i=x_{i-1}\exp({r_i})$ over the period from $i-1$ to $i$, then the
continuous, periodic rate of growth, $r_i$, is
\begin{equation}\label{cont}
  \text{continuously compounded return:}\qquad
  r_i = \log\left(\frac{x_i}{x_{i-1}}\right).
\end{equation}
Note that $r_i\eqref{cmpd}$ is just linear part of the series for $r_i\eqref{cont}$:
\[
  r_i\eqref{cont} = \sum_{k=1}^\infty \frac{(-1)^{k-1}}{k} r_i\eqref{cmpd}^k
  \quad \text{for} \quad -1 < r_i\eqref{cmpd} < 1.
\]

Tables \ref{cmpd39} and \ref{cont39} show the weekly and continuously
compounded weekly returns corresponding to the adjusted closing prices
in Table \ref{prices40a} or Table \ref{prices40b}. Either adjusted
closing price table produces the same weekly return table since
ratios of adjusted closing prices, $x_i/x_{i-1}$, depends only on the
security and not on the particular adjusted closing prices used.

The returns in Tables \ref{cmpd39} and \ref{cont39} are quite comparable
except that the weekly compounded returns \eqref{cmpd39} are
consistently greater than the continuously compounded returns
\eqref{cont39}. In fact a weekly compounded return can only equal a
continuously compounded return when $x_i = x_{i-1}$; then both returns
are zero.

\bigskip
We can now settle the question of which vector of security proportions
in the standard model corresponds to the investment portfolio
\texttt{PORTF}. We simply need to solve
\begin{equation}\tag{\ref{linear_hypothesis}}
  \mathbf{r}_P = \sum_{j=1}^n \mathbf{r}_j p_j = R \mathbf{p}.
\end{equation}
for $\mathbf{p}=[p_1,\ldots,p_n]^T$, using the periodic returns, $R$ and
$\mathbf{r}_P$, of Table \ref{cmpd39} or Table \ref{cont39}. Table
\ref{solvecmpd} shows the results---after the requirement $\sum_{j=1}^n
p_j=1$ has been enforced.

\begin{table}[th]
  \centering
  \caption{\label{solvecmpd}%
    Solution security proportions in \texttt{PORTF} (compound returns)
    }
  { \fontsize{10}{13}\selectfont
    \newcolumntype{e}{D{.}{.}{2}}
    \newcolumntype{f}{D{.}{.}{3}}
  \begin{tabular}{|c|fffff|e|}
    \hline\rule{0mm}{4mm}Table
    & \multicolumn{1}{c}{\texttt{IEF}} & \multicolumn{1}{c}{\texttt{IWB}}
    & \multicolumn{1}{c}{\texttt{IWM}} & \multicolumn{1}{c}{\texttt{EFA}}
    & \multicolumn{1}{c}{\texttt{EEM}} & \multicolumn{1}{|c|}{error} \\
    \hline \rule{0mm}{4mm}\ref{cmpd39}
    & 37.05\% & 38.15\% & 0.71\% & 24.08\% & 0.02\% & 3.3\% \\
    \hline \ref{cont39}
    & 37.22\% & 38.60\% & 0.32\% & 23.60\% & 0.25\% & 2.6\% \\
    \hline
  \end{tabular}
  }
\end{table}

The proportions in Table \ref{solvecmpd} are the least-squares solutions
of \eqref{linear_hypothesis} under the $\sum_{j=1}^n p_j=1$ constraint.
No matter that we started by investing \underline{exactly} 35\% of
\$100,000 in \texttt{IEF}, 40\% in \texttt{IWB}, and 25\% in
\texttt{EFA}, with nothing in \texttt{IWM} or \texttt{EEM}, at the close
of Thursday, 2009-12-31, there are no proportions $\mathbf{p}$ that can
make the linear hypothesis \eqref{linear_hypothesis} true with the
weekly returns of either Table \ref{cmpd39} or Table \ref{cont39}. With
the compound returns of Table \ref{cmpd39}, the theoretical portfolio
returns, $R\mathbf{p}$, can just get within 3.3\% of of the investment
portfolio returns, $\mathbf{r}_P$. With the continuous returns of Table
\ref{cont39}, $R\mathbf{p}$ can get within 2.6\%. In either case, the
least-squares-solution proportions of Table \ref{solvecmpd} are
ridiculous. There is no way you can start out by investing nothing in
\texttt{IWM} and \texttt{EEM} and end up with positive positions in both
of those securities.

Our conclusion is simple. Forget about the standard portfolio selection
model if you plan to use compound returns in the ex post version of the
model. The linear hypothesis \eqref{linear_hypothesis} will almost
certainly not hold---no matter how you choose the portfolio proportions
$\mathbf{p}$.


\section{Linear returns}
\label{linearreturns}

The problem with the periodically compounded periodic returns of the
last section,
\begin{equation}\tag{\ref{cmpd}}
  r_i = \frac{x_i}{x_{i-1}} - 1 = \frac{x_i - x_{i-1}}{x_{i-1}},
\end{equation}
is that the denominator, $x_{i-1}$, varies from one period to the next.
This non-linearity in the definition of the $r_i$ is incompatible with
the linear hypothesis of the standard model.

To remedy this problem one can choose a market-day-averaging-vector
$\bm{\alpha}$, as defined on page \pageref{marketdayavg}, and define
periodic returns $r^\alpha_i$ by
\begin{equation}\label{alphartn}
   \bm{\alpha}\text{-denominated linear return:}~\quad~
   r^\alpha_i = \frac{x_i - x_{i-1}}{\bm{\alpha}^T\mathbf{x}}.
\end{equation}
Then $r^\alpha_i$ is the percentage change in value of the security over
the $i-1$ to $i$ period relative to its $\bm{\alpha}$-average value.
Such \emph{$\bm{\alpha}$-denominated} or \emph{$\bm{\alpha}$-normalized}
linear returns are compatible with the linear hypothesis of the standard
model as we shall see.\\

\begin{rmk}
In this section and in the previous one we have used $i$ to index
successive periods, e.g. weeks. However the index in the dot product
$\bm{\alpha}^T\mathbf{x}$ of \eqref{alphartn} should not be restricted
to periodic values but should be allowed to range over all market days.
\end{rmk}

Tables \ref{linr39a} and \ref{linr39b} show [2009-12-31]-denominated and
$\alpha$-denominated linear returns corresponding to the adjusted
closing prices in Tables \ref{prices40a} and \ref{prices40b}. In fact the
returns in Tables \ref{linr39a} and \ref{linr39b} are just the
differences of the nomalized adjusted closing prices in Tables
\ref{prices40a} and \ref{prices40b}, respectively. Here again $\alpha$ is
the last-13-week market-day-averaging-vector of \eqref{alpha}.

Solving \begin{equation}\tag{\ref{linear_hypothesis}} \mathbf{r}_P =
\sum_{j=1}^n \mathbf{r}_j p_j = R \mathbf{p}. \end{equation} for
$\mathbf{p}$, with the $R$ and $\mathbf{r}_P$ of Tables \ref{linr39b}
and \ref{linr39b}, simply reproduces the notional portfolios
\eqref{portfa} and \eqref{portfb}:

\begin{table}[th]
  \centering
  \caption{\label{solvelnr}%
    Solution security proportions in \texttt{PORTF} (linear returns)
    }
  { \fontsize{10}{13}\selectfont
    \newcolumntype{f}{D{.}{.}{3}}
  \begin{tabular}{|c|fffff|c|}
    \hline\rule{0mm}{4mm}Table
    & \multicolumn{1}{c}{\texttt{IEF}} & \multicolumn{1}{c}{\texttt{IWB}}
    & \multicolumn{1}{c}{\texttt{IWM}} & \multicolumn{1}{c}{\texttt{EFA}}
    & \multicolumn{1}{c}{\texttt{EEM}} & \multicolumn{1}{|c|}{error} \\
    \hline \rule{0mm}{4mm}\ref{linr39a}
    & 35.00\% & 40.00\% & 0.00\% & 25.00\% & 0.00\% & $5.9\times 10^{-15}$ \\
    \hline \rule{0mm}{4mm}\ref{linr39b}
    & 35.65\% & 40.35\% & 0.00\% & 24.01\% & 0.00\% & $7.6\times 10^{-15}$ \\
    \hline
  \end{tabular}
  }
\end{table}

\bigskip
In general, the adjusted closing price equation
\begin{align}
  \mathbf{x}^\alpha_P &= X^\alpha \mathbf{p}^\alpha \tag{\ref{palpha}}\\
  \intertext{implies}
  \Delta\Theta\mathbf{x}^\alpha_P &=
    \Delta\Theta X^\alpha \mathbf{p}^\alpha, \label{matrixmult}\\
  \intertext{whence (dividing by the normalizing value, 100 in our case)}
  \mathbf{r}^\alpha_P &= R^\alpha \mathbf{p}^\alpha.
    \tag{\ref{linear_hypothesis}$\alpha$}
\end{align}
In equation \eqref{matrixmult}, $\Theta$ represents the
$(m+1)\times(M+1)$ submatrix $[\delta_{{i_k},i}]~ (k=0,1,\ldots,m;~
i=0,1,\ldots,M)$ of the $(M+1)\times(M+1)$ identity matrix that picks
out $m+1$ successive, periodic market days (rows) from a total of $M+1$
successive market days, and $\Delta$ is the $m\times(m+1)$ difference
matrix $[\delta_{ik}-\delta_{i,k+1}]~ (i=1,\ldots,m;~ k=0,1,\ldots,m)$.
(Here $\delta_{ij}$ is the Kronecker delta: $\delta_{ij}=1$ if $i=j$;
$\delta_{ij}=0$ otherwise.) Thus we see that $\bm{\alpha}$-normalized
linear returns and their corresponding notional portfolios \emph{always}
satisfy the linear hypothesis \eqref{linear_hypothesis} of the standard
mean-variance portfolio selection model.


\section{Some annualized statistics}
\label{annualizedstatistics}

We will close this paper with a brief discussion of some annualized
weekly return statistics for the last 39 weeks of 2010. These statistics
are based on the return Tables \ref{cmpd39}--\ref{linr39b} and uniform
weighting: $\omega_i=1/39~ (i=1,\ldots.39)$. To annualize expected
weekly return or variance of weekly return one simply multiplies by 52.
To annualize standard deviation of weekly return multiply by
$\sqrt{52}$.

Table \ref{returnstatistics} shows portfolio proportions $p$, mean
returns $e$, standard deviations of return $\sigma$, and
return-risk ratios $e\!/\!\sigma$ for the five exchange traded funds
and the investment portfolio \texttt{PORTF}.
Portfolio proportions make no sense for the compound returns; we have
filled these slots with question marks.

\begin{table}[th]
  \centering
  \caption{\label{returnstatistics}%
    Return statistics
    }
  { \fontsize{10}{13}\selectfont
    \newcolumntype{f}{D{.}{.}{3}}
  \begin{tabular}{|c|fffff|f|}
    \hline\rule{0mm}{4mm}Statistic
    & \multicolumn{1}{c}{\texttt{IEF}} & \multicolumn{1}{c}{\texttt{IWB}}
    & \multicolumn{1}{c}{\texttt{IWM}} & \multicolumn{1}{c}{\texttt{EFA}}
    & \multicolumn{1}{c}{\texttt{EEM}} & \multicolumn{1}{|c|}{\texttt{PORTF}} \\
    \hline
    \multicolumn{7}{|c|}{\rule{0mm}{4mm}Table \ref{cmpd39}: compounded weekly} \\
    \hline\rule{0mm}{4mm}$p$
    & \multicolumn{1}{c}{?} & \multicolumn{1}{c}{?} & \multicolumn{1}{c}{?}
    & \multicolumn{1}{c}{?} & \multicolumn{1}{c|}{?} & 100.00\% \\
    $e$ & 10.28\% & 13.30\% & 22.61\% & 8.94\% & 17.90\% & 10.45\% \\
    $\sigma$ & 6.53\% & 18.70\% & 25.66\% & 22.13\% & 23.98\% & 11.47\% \\
    $e\!/\!\sigma$ & 1.575 & 0.711 & 0.881 & 0.404 & 0.746 & 0.911 \\
    \hline
    \multicolumn{7}{|c|}{\rule{0mm}{4mm}Table \ref{cont39}: compounded continuously} \\
    \hline\rule{0mm}{4mm}$p$
    & \multicolumn{1}{c}{?} & \multicolumn{1}{c}{?} & \multicolumn{1}{c}{?}
    & \multicolumn{1}{c}{?} & \multicolumn{1}{c|}{?} & 100.00\% \\
    $e$ & 10.06\% & 11.52\% & 19.24\% & 6.46\% & 14.97\% & 9.78\% \\
    $\sigma$ & 6.54\% & 18.82\% & 25.92\% & 22.36\% & 24.13\% & 11.50\% \\
    $e\!/\!\sigma$ & 1.539 & 0.612 & 0.742 & 0.289 & 0.621 & 0.850 \\
    \hline
    \multicolumn{7}{|c|}{\rule{0mm}{4mm}Table \ref{linr39a}: [2009-12-31]-denominated linear} \\
    \hline\rule{0mm}{4mm}$p$
    & 35.00\% & 40.00\% & 0.00\% & 25.00\% & 0.00\% & 100.00\% \\
    $e$ & 10.60\% & 12.81\% & 22.74\% & 6.82\% & 16.50\% & 10.54\% \\
    $\sigma$ & 7.19\% & 19.09\% & 27.72\% & 21.30\% & 24.24\% & 11.83\% \\
    $e\!/\!\sigma$ & 1.474 & 0.671 & 0.820 & 0.320 & 0.681 & 0.890 \\
    \hline
    \multicolumn{7}{|c|}{\rule{0mm}{4mm}Table \ref{linr39b}: $\alpha$-denominated linear} \\
    \hline\rule{0mm}{4mm}$p$
    & 35.65\% & 40.35\% & 0.00\% & 24.01\% & 0.00\% & 100.00\% \\
    $e$ & 9.43\% & 11.51\% & 19.05\% & 6.43\% & 14.59\% & 9.55\% \\
    $\sigma$ & 6.40\% & 17.16\% & 23.23\% & 20.11\% & 21.42\% & 10.73\% \\
    $e\!/\!\sigma$ & 1.474 & 0.671 & 0.820 & 0.320 & 0.681 & 0.890 \\
    \hline
  \end{tabular}
  }
\end{table}

As noted earlier, continuously-compounded returns are ``always'' less
than periodically-compounded returns. The mean compound returns of Table
\ref{returnstatistics} reflect this relationship. Expected-values and
standard-deviations of linear returns depend on their normalizations.
This is clear in Table \ref{returnstatistics}. However the linear
return-risk ratio $e\!/\!\sigma$ is independent of normalization:
normalizing factors cancel out, top and bottom. Correlations of linear
returns are independent of normalization for the same reason.

\begin{table}[th]
  \centering
  \caption{\label{correlation}%
    Correlation of linear returns
    }
  { \fontsize{10}{13}\selectfont
    \newcolumntype{f}{D{.}{.}{3}}
  $
  \begin{tabular}{|c|fffff|}
    \hline\rule{0mm}{4mm}\text{fund}
    & \multicolumn{1}{c}{\texttt{~IEF}} & \multicolumn{1}{c}{\texttt{~IWB}}
    & \multicolumn{1}{c}{\texttt{~IWM}} & \multicolumn{1}{c}{\texttt{~EFA}}
    & \multicolumn{1}{c|}{\texttt{~EEM}} \\
    \hline\multicolumn{1}{|c|}{\rule{0mm}{4mm}\texttt{IEF}}
      & 1.000 & -0.470 & -0.497 & -0.334 & -0.296 \\
    \multicolumn{1}{|c|}{\texttt{IWB}}
      & -0.470 &  1.000 &  0.948 &  0.911 &  0.887 \\
    \multicolumn{1}{|c|}{\texttt{IWM}}
      & -0.497 &  0.948 &  1.000 &  0.817 &  0.829 \\
    \multicolumn{1}{|c|}{\texttt{EFA}}
      & -0.334 &  0.911 &  0.817 &  1.000 &  0.904 \\
    \multicolumn{1}{|c|}{\texttt{EEM}}
      & -0.296 &  0.887 &  0.829 &  0.904 &  1.000 \\
    \hline
  \end{tabular}
  $
  }
\end{table}

We could compute the correlation coefficients $c_{jk}$ corresponding to
the compound returns of Tables \ref{cmpd39} and \ref{cont39}. Then the
compound covariance coefficients would be defined by
\begin{equation}\label{covariance2}
  v_{jk}=c_{jk}\sigma_j\sigma_k
  \quad (j, k = 1,\ldots n).
\end{equation}
But what is the point of computing such or $c_{jk}$ and $v_{jk}$ when
the $p_j$ and $p_k$ in
\begin{equation}\tag{\ref{variance}}
  v_P = \sum_{j=1}^n\sum_{k=1}^n v_{jk} p_j p_k,
\end{equation}
don't even exist?

\bigskip
We will close this section and the paper by showing that the
$e\!/\!\sigma$ ratios and the correlation coefficients of normalized
linear returns are independent of the normalization.

For the $e\!/\!\sigma$ case start with an adjusted-closing-price vector
$\mathbf{x}$ and a market-day-averaging-vector $\bm{\alpha}$. Let
$e^\alpha$ and $\sigma^\alpha$ denote the expected value and standard
deviation of the periodic return vector $\mathbf{r}^\alpha$. Then
\[
  e^\alpha = \bm{\omega}^T\mathbf{r}^\alpha
    = \bm{\omega}^T(\Delta\Theta\mathbf{x})
      \left(\frac{1}{\bm{\alpha}^T\mathbf{x}}\right),
\]
and
\[
  \sigma^\alpha = \|\mathbf{r}^\alpha-\mathbf{1}_m e^\alpha\|_\omega
    = \|(I_m - \mathbf{1}_m\bm{\omega}^T)\mathbf{r}^\alpha\|_\omega
    = \|(I_m - \mathbf{1}_m\bm{\omega}^T)(\Delta\Theta\mathbf{x})\|_\omega
      \left(\frac{1}{\bm{\alpha}^T\mathbf{x}}\right),
\]
where $\|\mathbf{z}\|_\omega=\sqrt{\sum_{i=1}^m\omega_iz_i^2}$ and
$\Delta$ and $\Theta$ are the matrices in of \eqref{matrixmult}. The
factors $1/(\bm{\alpha}^T\mathbf{x})$ cancel in the $e\!/\!\sigma$ ratio
so that
\begin{equation}\label{returnriskratio}
  \frac{e}{\sigma} = \frac{e^\alpha}{\sigma^\alpha}
    = \frac{\bm{\omega}^T(\Delta\Theta\mathbf{x})}
      {\|(I_m - \mathbf{1}_m\bm{\omega}^T)(\Delta\Theta\mathbf{x})\|_\omega}
\end{equation}
does not depend on $\bm{\alpha}$ at all. Multiplying $\mathbf{x}$ by a
positive number has no effect on the right side of
\eqref{returnriskratio}; so the $e\!/\!\sigma$ ratio depends only on the
security, not on the particular adjusted closing prices that describe
its growth. Of course, the $e\!/\!\sigma$ ratio also depends on the
weight system $\bm{\omega}$.

The ``$\alpha$'' correlation coefficient $c^\alpha_{jk}$ is defined by
\[
  c^\alpha_{jk} = \frac{(\mathbf{z}^\alpha_j)^T
    \diag(\bm{\omega})\mathbf{z}^\alpha_k}
    {\|\mathbf{z}^\alpha_j\|_\omega\|\mathbf{z}^\alpha_k\|_\omega}
    \quad \text{with} \quad
    \mathbf{z}^\alpha_* = \mathbf{r}^\alpha_*-\mathbf{1}_m e^\alpha_*. 
\]
Following the pattern of the $e\!/\!\sigma$ demonstration we compute
\begin{align}
  c_{jk} &= c^\alpha_{jk} \notag \\
    &= \frac{
      (\Delta\Theta\mathbf{x}_j)^T(I_m-\mathbf{1}_m\bm{\omega}^T)^T
      \diag(\bm{\omega})
      (I_m-\mathbf{1}_m\bm{\omega}^T)(\Delta\Theta\mathbf{x}_k)
      }{
      \|(I_m - \mathbf{1}_m\bm{\omega}^T)(\Delta\Theta\mathbf{x}_j)\|_\omega
      \cdot
      \|(I_m - \mathbf{1}_m\bm{\omega}^T)(\Delta\Theta\mathbf{x}_k)\|_\omega
      } \notag \\
    &= \frac{
      (\Delta\Theta\mathbf{x}_j)^T
      \left[\diag(\bm{\omega})-\bm{\omega}\bm{\omega}^T\right]
      (\Delta\Theta\mathbf{x}_k)
      }{
      \|(I_m - \mathbf{1}_m\bm{\omega}^T)(\Delta\Theta\mathbf{x}_j)\|_\omega
      \cdot
      \|(I_m - \mathbf{1}_m\bm{\omega}^T)(\Delta\Theta\mathbf{x}_k)\|_\omega
      }, \label{correlationexpression}
\end{align}
with the normalizing factors, $\bm{\alpha}^T\mathbf{x}_j$ and
$\bm{\alpha}^T\mathbf{x}_k$, canceling out. Since the expression
\eqref{correlationexpression} is unchanged when $\mathbf{x}_j$ and
$\mathbf{x}_k$ are multiplied by positive numbers, $c_{jk}$ is
independent of the particular adjusted closing prices representing the
$j$ and $k$ securities.
\qed


\appendix
\numberwithin{table}{section}
\numberwithin{equation}{section}
\numberwithin{figure}{section}

\section{Closing portfolios}
\label{appendix:closingportf40}

\renewcommand\thempfootnote{\fnsymbol{mpfootnote}}
\vspace*{-2.0ex}
\begin{center}
\begin{minipage}{4.85in}
\begin{table}[H]
\centering
\caption{\label{portf40}
   Week-closing portfolios for the last 40 weeks of 2010
}
{ \fontsize{10}{13}\selectfont
  \newcolumntype{.}{D{.}{.}{3}}
\begin{tabular}{|l|..c.c|c|} \hline
  \multicolumn{1}{|c|}{\rule{0mm}{4mm}\textrm{Friday}}
  & \multicolumn{1}{c}{\texttt{IEF}} & \multicolumn{1}{c}{\texttt{IWB}}
  & \texttt{IWM} & \multicolumn{1}{c}{\texttt{EFA}}
  & \texttt{EEM} & \texttt{PORTF} \\
\hline
  2010-04-02\mpfootnotemark[1]
             & 34.18\% & 41.01\% & 0\% & 24.81\% & 0\% & 100\% \\
  2010-04-09 & 33.94\% & 41.35\% & 0\% & 24.71\% & 0\% & 100\% \\
  2010-04-16 & 34.23\% & 41.28\% & 0\% & 24.50\% & 0\% & 100\% \\
  2010-04-23 & 33.88\% & 41.89\% & 0\% & 24.23\% & 0\% & 100\% \\
  2010-04-30 & 34.81\% & 41.49\% & 0\% & 23.69\% & 0\% & 100\% \\
  2010-05-07 & 37.05\% & 40.51\% & 0\% & 22.44\% & 0\% & 100\% \\
  2010-05-14 & 36.54\% & 40.99\% & 0\% & 22.47\% & 0\% & 100\% \\
  2010-05-21 & 37.82\% & 39.92\% & 0\% & 22.26\% & 0\% & 100\% \\
  2010-05-28 & 37.70\% & 40.17\% & 0\% & 22.13\% & 0\% & 100\% \\
  2010-06-04 & 38.49\% & 39.76\% & 0\% & 21.75\% & 0\% & 100\% \\
  2010-06-11 & 37.72\% & 40.00\% & 0\% & 22.28\% & 0\% & 100\% \\
  2010-06-18 & 37.12\% & 40.22\% & 0\% & 22.66\% & 0\% & 100\% \\
  2010-06-25 & 38.03\% & 39.39\% & 0\% & 22.58\% & 0\% & 100\% \\
  2010-07-02 & 39.44\% & 38.22\% & 0\% & 22.34\% & 0\% & 100\% \\
  2010-07-09 & 37.95\% & 39.07\% & 0\% & 22.98\% & 0\% & 100\% \\
  2010-07-16 & 38.52\% & 38.66\% & 0\% & 22.81\% & 0\% & 100\% \\
  2010-07-23 & 37.55\% & 39.22\% & 0\% & 23.23\% & 0\% & 100\% \\
  2010-07-30 & 37.71\% & 39.02\% & 0\% & 23.27\% & 0\% & 100\% \\
  2010-08-06 & 37.34\% & 39.00\% & 0\% & 23.66\% & 0\% & 100\% \\
  2010-08-13 & 38.71\% & 38.44\% & 0\% & 22.86\% & 0\% & 100\% \\
  2010-08-20 & 38.91\% & 38.35\% & 0\% & 22.74\% & 0\% & 100\% \\
  2010-08-27 & 38.91\% & 38.20\% & 0\% & 22.89\% & 0\% & 100\% \\
  2010-09-03 & 37.89\% & 38.83\% & 0\% & 23.28\% & 0\% & 100\% \\
  2010-09-10 & 37.64\% & 38.96\% & 0\% & 23.40\% & 0\% & 100\% \\
  2010-09-17 & 37.41\% & 39.10\% & 0\% & 23.49\% & 0\% & 100\% \\
  2010-09-24 & 37.07\% & 39.14\% & 0\% & 23.79\% & 0\% & 100\% \\
  2010-10-01 & 37.24\% & 39.02\% & 0\% & 23.75\% & 0\% & 100\% \\
  2010-10-08 & 37.05\% & 39.00\% & 0\% & 23.95\% & 0\% & 100\% \\
  2010-10-15 & 36.52\% & 39.31\% & 0\% & 24.17\% & 0\% & 100\% \\
  2010-10-22 & 36.53\% & 39.45\% & 0\% & 24.02\% & 0\% & 100\% \\
  2010-10-29 & 36.49\% & 39.57\% & 0\% & 23.93\% & 0\% & 100\% \\
  2010-11-05 & 35.86\% & 39.92\% & 0\% & 24.21\% & 0\% & 100\% \\
  2010-11-12 & 36.09\% & 39.94\% & 0\% & 23.97\% & 0\% & 100\% \\
  2010-11-19 & 35.82\% & 40.03\% & 0\% & 24.14\% & 0\% & 100\% \\
  2010-11-26 & 36.34\% & 40.21\% & 0\% & 23.45\% & 0\% & 100\% \\
  2010-12-03 & 35.29\% & 40.81\% & 0\% & 23.89\% & 0\% & 100\% \\
  2010-12-10 & 34.51\% & 41.43\% & 0\% & 24.06\% & 0\% & 100\% \\
  2010-12-17 & 34.52\% & 41.51\% & 0\% & 23.97\% & 0\% & 100\% \\
  2010-12-24\mpfootnotemark[1]
             & 34.18\% & 41.71\% & 0\% & 24.12\% & 0\% & 100\% \\
  2010-12-31 & 34.25\% & 41.54\% & 0\% & 24.20\% & 0\% & 100\% \\
\hline
\end{tabular}
}
\end{table}
\footnotetext[1]{Friday market-holiday $=$
  Thursday closing portfolio}
\end{minipage}
\end{center}

\newpage
\section{Adjusted closing prices}
\label{appendix:prices40}

\vspace*{-5.0ex}
\begin{center}
\begin{minipage}{5.55in}
\begin{table}[H]
\centering
\caption{\label{prices40a}
   [2009-12-31]-normalized adjusted closing prices for the last 40 weeks of 2010\\[0.5ex]
   $ \texttt{PORTF} = 35.00\%\times\texttt{IEF} + 40.00\%\times\texttt{IWB}
   + 0\%\times\texttt{IWM} + 25.00\%\times\texttt{EFA} + 0\%\times\texttt{EEM}
   $
}
{ \fontsize{10}{13}\selectfont
  \newcolumntype{.}{D{.}{.}{3}}
\begin{tabular}{|l|.....|.|} \hline
  \multicolumn{1}{|c|}{\rule{0mm}{4mm}\textrm{Friday}}
  & \multicolumn{1}{c}{\texttt{IEF}} & \multicolumn{1}{c}{\texttt{IWB}}
  & \multicolumn{1}{c}{\texttt{IWM}} & \multicolumn{1}{c}{\texttt{EFA}}
  & \multicolumn{1}{c}{\texttt{EEM}}
  & \multicolumn{1}{|c|}{\texttt{PORTF}} \\
\hline
  2010-04-02\mpfootnotemark[1]
             & 101.413 & 106.448 & 109.866 & 103.057 & 104.145 & 103.838 \\
  2010-04-09 & 101.424 & 108.118 & 112.788 & 103.383 & 105.494 & 104.591 \\
  2010-04-16 & 102.403 & 108.053 & 114.731 & 102.605 & 102.313 & 104.714 \\
  2010-04-23 & 102.005 & 110.345 & 119.017 & 102.135 & 103.759 & 105.374 \\
  2010-04-30 & 103.303 & 107.725 & 115.036 &  98.426 & 101.325 & 103.853 \\
  2010-05-07 & 104.986 & 100.440 & 104.937 &  89.020 &  92.024 &  99.176 \\
  2010-05-14 & 105.008 & 103.076 & 111.680 &  90.431 &  95.157 & 100.591 \\
  2010-05-21 & 106.722 &  98.590 & 104.471 &  87.952 &  89.976 &  98.777 \\
  2010-05-28 & 106.333 &  99.146 & 106.366 &  87.410 &  91.807 &  98.727 \\
  2010-06-04 & 107.181 &  96.871 & 102.047 &  84.805 &  89.639 &  97.463 \\
  2010-06-11 & 106.974 &  99.277 & 104.263 &  88.477 &  93.398 &  99.271 \\
  2010-06-18 & 107.135 & 101.586 & 107.249 &  91.552 &  96.193 & 101.020 \\
  2010-06-25 & 108.177 &  98.033 & 103.765 &  89.914 &  95.640 &  99.554 \\
  2010-07-02 & 109.331 &  92.703 &  96.359 &  86.674 &  91.565 &  97.016 \\
  2010-07-09 & 108.700 &  97.920 & 101.385 &  92.141 &  96.950 & 100.248 \\
  2010-07-16 & 110.078 &  96.669 &  98.372 &  91.257 &  93.748 & 100.009 \\
  2010-07-23 & 109.641 & 100.208 & 104.671 &  94.957 &  99.812 & 102.197 \\
  2010-07-30 & 110.571 & 100.109 & 104.735 &  95.546 & 100.419 & 102.630 \\
  2010-08-06 & 111.542 & 101.936 & 104.928 &  98.933 & 102.068 & 104.547 \\
  2010-08-13 & 112.912 &  98.101 &  98.372 &  93.337 &  98.672 & 102.094 \\
  2010-08-20 & 113.177 &  97.608 &  98.501 &  92.601 &  99.497 & 101.805 \\
  2010-08-27 & 112.912 &  96.982 &  99.307 &  92.987 &  98.211 & 101.559 \\
  2010-09-03 & 112.340 & 100.734 & 103.624 &  96.613 & 101.947 & 103.766 \\
  2010-09-10 & 111.705 & 101.162 & 102.641 &  97.202 & 102.505 & 103.862 \\
  2010-09-17 & 112.271 & 102.693 & 105.041 &  98.712 & 104.348 & 105.050 \\
  2010-09-24 & 113.437 & 104.798 & 108.180 & 101.933 & 107.016 & 107.105 \\
  2010-10-01 & 114.341 & 104.831 & 109.585 & 102.098 & 110.194 & 107.476 \\
  2010-10-08 & 115.672 & 106.551 & 111.878 & 104.675 & 112.134 & 109.274 \\
  2010-10-15 & 114.295 & 107.658 & 113.509 & 105.890 & 113.323 & 109.539 \\
  2010-10-22 & 114.561 & 108.237 & 113.558 & 105.448 & 111.649 & 109.753 \\
  2010-10-29 & 114.272 & 108.435 & 113.525 & 104.933 & 111.867 & 109.602 \\
  2010-11-05 & 115.274 & 112.287 & 119.129 & 108.964 & 117.616 & 112.502 \\
  2010-11-12 & 113.686 & 110.072 & 116.319 & 105.724 & 112.571 & 110.250 \\
  2010-11-19 & 112.700 & 110.204 & 116.997 & 106.332 & 112.813 & 110.110 \\
  2010-11-26 & 112.804 & 109.196 & 118.241 & 101.914 & 108.666 & 108.638 \\
  2010-12-03 & 111.489 & 112.800 & 122.197 & 105.669 & 114.341 & 110.558 \\
  2010-12-10 & 108.687 & 114.189 & 125.556 & 106.074 & 113.007 & 110.235 \\
  2010-12-17 & 109.012 & 114.701 & 125.992 & 105.964 & 112.546 & 110.526 \\
  2010-12-24\mpfootnotemark[1]
             & 108.547 & 115.906 & 127.698 & 107.240 & 113.942 & 111.164 \\
  2010-12-31 & 109.360 & 116.056 & 126.919 & 108.169 & 116.523 & 111.741 \\
\hline
\end{tabular}
}
\end{table}
\footnotetext[1]{Friday market-holiday $=$
  Thursday adjusted closing prices}
\end{minipage}
\end{center}

\newpage
\begin{center}
\begin{minipage}{5.55in}
\begin{table}[H]
\centering
\caption{\label{prices40b}
   $\alpha$-normalized adjusted closing prices for the last 40 weeks of 2010\\[0.5ex]
   $ \texttt{PORTF} = 35.65\%\times\texttt{IEF} + 40.35\%\times\texttt{IWB}
   + 0\%\times\texttt{IWM} + 24.01\%\times\texttt{EFA} + 0\%\times\texttt{EEM}
   $
}
{ \fontsize{10}{13}\selectfont
  \newcolumntype{.}{D{.}{.}{3}}
\begin{tabular}{|l|.....|.|} \hline
  \multicolumn{1}{|c|}{\rule{0mm}{4mm}\textrm{Friday}}
  & \multicolumn{1}{c}{\texttt{IEF}} & \multicolumn{1}{c}{\texttt{IWB}}
  & \multicolumn{1}{c}{\texttt{IWM}} & \multicolumn{1}{c}{\texttt{EFA}}
  & \multicolumn{1}{c}{\texttt{EEM}}
  & \multicolumn{1}{|c|}{\texttt{PORTF}} \\
\hline
  2010-04-02\mpfootnotemark[1]
             &  90.277 &  95.681 &  92.056 &  97.294 &  92.039 &  94.142 \\
  2010-04-09 &  90.287 &  97.182 &  94.504 &  97.602 &  93.231 &  94.825 \\
  2010-04-16 &  91.158 &  97.123 &  96.132 &  96.868 &  90.419 &  94.936 \\
  2010-04-23 &  90.804 &  99.184 &  99.723 &  96.424 &  91.697 &  95.534 \\
  2010-04-30 &  91.960 &  96.829 &  96.387 &  92.922 &  89.546 &  94.155 \\
  2010-05-07 &  93.458 &  90.281 &  87.926 &  84.042 &  81.327 &  89.915 \\
  2010-05-14 &  93.477 &  92.650 &  93.575 &  85.374 &  84.095 &  91.198 \\
  2010-05-21 &  95.003 &  88.618 &  87.535 &  83.034 &  79.517 &  89.553 \\
  2010-05-28 &  94.657 &  89.117 &  89.123 &  82.522 &  81.135 &  89.509 \\
  2010-06-04 &  95.412 &  87.073 &  85.504 &  80.063 &  79.219 &  88.362 \\
  2010-06-11 &  95.227 &  89.235 &  87.361 &  83.530 &  82.541 &  90.001 \\
  2010-06-18 &  95.371 &  91.311 &  89.863 &  86.433 &  85.011 &  91.587 \\
  2010-06-25 &  96.298 &  88.117 &  86.944 &  84.886 &  84.522 &  90.258 \\
  2010-07-02 &  97.326 &  83.326 &  80.738 &  81.828 &  80.921 &  87.957 \\
  2010-07-09 &  96.764 &  88.015 &  84.949 &  86.989 &  85.680 &  90.887 \\
  2010-07-16 &  97.991 &  86.891 &  82.425 &  86.154 &  82.850 &  90.671 \\
  2010-07-23 &  97.602 &  90.072 &  87.703 &  89.647 &  88.209 &  92.654 \\
  2010-07-30 &  98.429 &  89.983 &  87.756 &  90.203 &  88.746 &  93.047 \\
  2010-08-06 &  99.294 &  91.625 &  87.918 &  93.401 &  90.203 &  94.785 \\
  2010-08-13 & 100.513 &  88.178 &  82.425 &  88.118 &  87.202 &  92.561 \\
  2010-08-20 & 100.749 &  87.735 &  82.533 &  87.423 &  87.931 &  92.299 \\
  2010-08-27 & 100.513 &  87.172 &  83.208 &  87.788 &  86.794 &  92.076 \\
  2010-09-03 & 100.004 &  90.545 &  86.825 &  91.211 &  90.096 &  94.077 \\
  2010-09-10 &  99.439 &  90.929 &  86.002 &  91.767 &  90.589 &  94.164 \\
  2010-09-17 &  99.943 &  92.306 &  88.013 &  93.192 &  92.218 &  95.241 \\
  2010-09-24 & 100.981 &  94.198 &  90.643 &  96.233 &  94.576 &  97.104 \\
  2010-10-01 & 101.785 &  94.227 &  91.820 &  96.389 &  97.384 &  97.441 \\
  2010-10-08 & 102.970 &  95.773 &  93.741 &  98.822 &  99.099 &  99.071 \\
  2010-10-15 & 101.745 &  96.768 &  95.108 &  99.969 & 100.150 &  99.311 \\
  2010-10-22 & 101.981 &  97.289 &  95.149 &  99.552 &  98.670 &  99.505 \\
  2010-10-29 & 101.724 &  97.467 &  95.121 &  99.066 &  98.863 &  99.368 \\
  2010-11-05 & 102.616 & 100.929 &  99.817 & 102.871 & 103.944 & 101.997 \\
  2010-11-12 & 101.202 &  98.938 &  97.462 &  99.812 &  99.485 &  99.955 \\
  2010-11-19 & 100.325 &  99.057 &  98.031 & 100.386 &  99.699 &  99.828 \\
  2010-11-26 & 100.417 &  98.151 &  99.073 &  96.215 &  96.034 &  98.494 \\
  2010-12-03 &  99.247 & 101.390 & 102.388 &  99.760 & 101.049 & 100.235 \\
  2010-12-10 &  96.752 & 102.639 & 105.202 & 100.143 &  99.870 &  99.941 \\
  2010-12-17 &  97.042 & 103.099 & 105.567 & 100.039 &  99.463 & 100.205 \\
  2010-12-24\mpfootnotemark[1]
             &  96.628 & 104.182 & 106.997 & 101.244 & 100.697 & 100.784 \\
  2010-12-31 &  97.351 & 104.317 & 106.344 & 102.121 & 102.978 & 101.307 \\
\hline
\end{tabular}
}
\end{table}
\footnotetext[1]{Friday market-holiday $=$
  Thursday adjusted closing prices}
\end{minipage}
\end{center}

\newpage
\section{Compound returns}
\label{appendix:compoundreturns39}

\begin{center}
\begin{minipage}{6.00in}
\begin{table}[H]
\centering
\caption{\label{cmpd39}
   Compound weekly returns (\%) for the last 39 weeks of 2010.
}
{ \fontsize{10}{13}\selectfont
  \newcolumntype{.}{D{.}{.}{3}}
$
\begin{array}{|l|.....|.|} \hline
  \multicolumn{1}{|c|}{\rule{0mm}{4mm}\textrm{Friday}}
  & \multicolumn{1}{c}{\texttt{~IEF}} & \multicolumn{1}{c}{\texttt{~IWB}}
  & \multicolumn{1}{c}{\texttt{~IWM}} & \multicolumn{1}{c}{\texttt{~~EFA}}
  & \multicolumn{1}{c}{\texttt{~EEM}} & \multicolumn{1}{|c|}{\texttt{~PORTF}} \\
\hline
  \text{2010-04-09} &   0.011 &   1.569 &   2.660 &   0.316 &   1.295 &   0.726 \\
  \text{2010-04-16} &   0.965 &  -0.060 &   1.723 &  -0.753 &  -3.015 &   0.117 \\
  \text{2010-04-23} &  -0.389 &   2.121 &   3.736 &  -0.458 &   1.413 &   0.630 \\
  \text{2010-04-30} &   1.272 &  -2.374 &  -3.345 &  -3.631 &  -2.346 &  -1.443 \\
  \text{2010-05-07} &   1.629 &  -6.763 &  -8.779 &  -9.556 &  -9.179 &  -4.503 \\
  \text{2010-05-14} &   0.021 &   2.624 &   6.426 &   1.585 &   3.405 &   1.427 \\
  \text{2010-05-21} &   1.632 &  -4.352 &  -6.455 &  -2.741 &  -5.445 &  -1.804 \\
  \text{2010-05-28} &  -0.364 &   0.564 &   1.814 &  -0.616 &   2.035 &  -0.050 \\
  \text{2010-06-04} &   0.797 &  -2.295 &  -4.061 &  -2.980 &  -2.361 &  -1.281 \\
  \text{2010-06-11} &  -0.193 &   2.484 &   2.172 &   4.330 &   4.193 &   1.855 \\
  \text{2010-06-18} &   0.151 &   2.326 &   2.864 &   3.475 &   2.993 &   1.762 \\
  \text{2010-06-25} &   0.973 &  -3.498 &  -3.249 &  -1.789 &  -0.575 &  -1.451 \\
  \text{2010-07-02} &   1.067 &  -5.437 &  -7.137 &  -3.603 &  -4.261 &  -2.549 \\
  \text{2010-07-09} &  -0.577 &   5.628 &   5.216 &   6.308 &   5.881 &   3.332 \\
  \text{2010-07-16} &   1.268 &  -1.278 &  -2.972 &  -0.959 &  -3.303 &  -0.239 \\
  \text{2010-07-23} &  -0.397 &   3.661 &   6.403 &   4.054 &   6.468 &   2.187 \\
  \text{2010-07-30} &   0.848 &  -0.099 &   0.061 &   0.620 &   0.608 &   0.424 \\
  \text{2010-08-06} &   0.878 &   1.825 &   0.184 &   3.545 &   1.642 &   1.868 \\
  \text{2010-08-13} &   1.228 &  -3.762 &  -6.248 &  -5.656 &  -3.327 &  -2.347 \\
  \text{2010-08-20} &   0.235 &  -0.503 &   0.131 &  -0.789 &   0.836 &  -0.283 \\
  \text{2010-08-27} &  -0.234 &  -0.641 &   0.818 &   0.417 &  -1.293 &  -0.242 \\
  \text{2010-09-03} &  -0.507 &   3.869 &   4.347 &   3.899 &   3.804 &   2.173 \\
  \text{2010-09-10} &  -0.565 &   0.425 &  -0.949 &   0.610 &   0.547 &   0.093 \\
  \text{2010-09-17} &   0.507 &   1.513 &   2.338 &   1.553 &   1.798 &   1.144 \\
  \text{2010-09-24} &   1.039 &   2.050 &   2.988 &   3.263 &   2.557 &   1.957 \\
  \text{2010-10-01} &   0.797 &   0.031 &   1.299 &   0.162 &   2.970 &   0.346 \\
  \text{2010-10-08} &   1.164 &   1.641 &   2.092 &   2.524 &   1.761 &   1.673 \\
  \text{2010-10-15} &  -1.190 &   1.039 &   1.458 &   1.161 &   1.060 &   0.242 \\
  \text{2010-10-22} &   0.233 &   0.538 &   0.043 &  -0.417 &  -1.477 &   0.196 \\
  \text{2010-10-29} &  -0.252 &   0.183 &  -0.029 &  -0.488 &   0.195 &  -0.137 \\
  \text{2010-11-05} &   0.877 &   3.552 &   4.936 &   3.841 &   5.139 &   2.645 \\
  \text{2010-11-12} &  -1.378 &  -1.973 &  -2.359 &  -2.973 &  -4.289 &  -2.002 \\
  \text{2010-11-19} &  -0.867 &   0.120 &   0.583 &   0.575 &   0.215 &  -0.127 \\
  \text{2010-11-26} &   0.092 &  -0.915 &   1.063 &  -4.155 &  -3.676 &  -1.336 \\
  \text{2010-12-03} &  -1.166 &   3.300 &   3.346 &   3.684 &   5.222 &   1.767 \\
  \text{2010-12-10} &  -2.513 &   1.231 &   2.749 &   0.383 &  -1.167 &  -0.293 \\
  \text{2010-12-17} &   0.299 &   0.448 &   0.347 &  -0.104 &  -0.408 &   0.264 \\
  \text{2010-12-24} &  -0.427 &   1.051 &   1.354 &   1.204 &   1.240 &   0.577 \\
  \text{2010-12-31} &   0.749 &   0.129 &  -0.610 &   0.866 &   2.265 &   0.519 \\
\hline
\end{array}
$
}
\end{table}
\end{minipage}
\end{center}

\newpage
\begin{center}
\begin{minipage}{6.00in}
\begin{table}[H]
\centering
\caption{\label{cont39}
   Continuous weekly returns (\%) for the last 39 weeks of 2010.
}
{ \fontsize{10}{13}\selectfont
  \newcolumntype{.}{D{.}{.}{3}}
$
\begin{array}{|l|.....|.|} \hline
  \multicolumn{1}{|c|}{\rule{0mm}{4mm}\textrm{Friday}}
  & \multicolumn{1}{c}{\texttt{~IEF}} & \multicolumn{1}{c}{\texttt{~IWB}}
  & \multicolumn{1}{c}{\texttt{~IWM}} & \multicolumn{1}{c}{\texttt{~EFA}}
  & \multicolumn{1}{c}{\texttt{~EEM}} & \multicolumn{1}{|c|}{\texttt{~PORTF}} \\
\hline
  \text{2010-04-09} &   0.011 &   1.557 &   2.625 &   0.316 &   1.287 &   0.723 \\
  \text{2010-04-16} &   0.961 &  -0.060 &   1.708 &  -0.755 &  -3.062 &   0.117 \\
  \text{2010-04-23} &  -0.389 &   2.099 &   3.668 &  -0.459 &   1.403 &   0.628 \\
  \text{2010-04-30} &   1.264 &  -2.403 &  -3.402 &  -3.699 &  -2.374 &  -1.454 \\
  \text{2010-05-07} &   1.616 &  -7.002 &  -9.188 & -10.044 &  -9.628 &  -4.608 \\
  \text{2010-05-14} &   0.021 &   2.591 &   6.228 &   1.573 &   3.348 &   1.417 \\
  \text{2010-05-21} &   1.619 &  -4.450 &  -6.673 &  -2.780 &  -5.599 &  -1.820 \\
  \text{2010-05-28} &  -0.365 &   0.562 &   1.798 &  -0.618 &   2.015 &  -0.050 \\
  \text{2010-06-04} &   0.794 &  -2.321 &  -4.145 &  -3.026 &  -2.390 &  -1.289 \\
  \text{2010-06-11} &  -0.193 &   2.453 &   2.148 &   4.239 &   4.108 &   1.838 \\
  \text{2010-06-18} &   0.150 &   2.299 &   2.824 &   3.416 &   2.949 &   1.746 \\
  \text{2010-06-25} &   0.968 &  -3.560 &  -3.302 &  -1.805 &  -0.577 &  -1.462 \\
  \text{2010-07-02} &   1.061 &  -5.590 &  -7.405 &  -3.670 &  -4.354 &  -2.583 \\
  \text{2010-07-09} &  -0.579 &   5.475 &   5.084 &   6.117 &   5.715 &   3.278 \\
  \text{2010-07-16} &   1.260 &  -1.286 &  -3.017 &  -0.964 &  -3.359 &  -0.239 \\
  \text{2010-07-23} &  -0.398 &   3.596 &   6.207 &   3.974 &   6.268 &   2.164 \\
  \text{2010-07-30} &   0.845 &  -0.099 &   0.061 &   0.618 &   0.606 &   0.423 \\
  \text{2010-08-06} &   0.874 &   1.809 &   0.184 &   3.484 &   1.629 &   1.851 \\
  \text{2010-08-13} &   1.221 &  -3.835 &  -6.452 &  -5.823 &  -3.384 &  -2.375 \\
  \text{2010-08-20} &   0.234 &  -0.504 &   0.131 &  -0.792 &   0.833 &  -0.283 \\
  \text{2010-08-27} &  -0.234 &  -0.643 &   0.815 &   0.416 &  -1.301 &  -0.243 \\
  \text{2010-09-03} &  -0.508 &   3.796 &   4.255 &   3.825 &   3.733 &   2.150 \\
  \text{2010-09-10} &  -0.567 &   0.424 &  -0.953 &   0.608 &   0.546 &   0.093 \\
  \text{2010-09-17} &   0.505 &   1.502 &   2.311 &   1.542 &   1.782 &   1.137 \\
  \text{2010-09-24} &   1.033 &   2.029 &   2.945 &   3.211 &   2.525 &   1.938 \\
  \text{2010-10-01} &   0.794 &   0.031 &   1.290 &   0.162 &   2.926 &   0.346 \\
  \text{2010-10-08} &   1.157 &   1.627 &   2.071 &   2.493 &   1.745 &   1.659 \\
  \text{2010-10-15} &  -1.198 &   1.034 &   1.447 &   1.154 &   1.055 &   0.242 \\
  \text{2010-10-22} &   0.232 &   0.536 &   0.043 &  -0.418 &  -1.488 &   0.195 \\
  \text{2010-10-29} &  -0.253 &   0.183 &  -0.029 &  -0.490 &   0.195 &  -0.137 \\
  \text{2010-11-05} &   0.873 &   3.491 &   4.818 &   3.770 &   5.011 &   2.611 \\
  \text{2010-11-12} &  -1.387 &  -1.992 &  -2.387 &  -3.019 &  -4.384 &  -2.022 \\
  \text{2010-11-19} &  -0.871 &   0.120 &   0.581 &   0.573 &   0.215 &  -0.127 \\
  \text{2010-11-26} &   0.092 &  -0.919 &   1.058 &  -4.244 &  -3.745 &  -1.345 \\
  \text{2010-12-03} &  -1.173 &   3.247 &   3.291 &   3.618 &   5.091 &   1.752 \\
  \text{2010-12-10} &  -2.545 &   1.224 &   2.712 &   0.383 &  -1.174 &  -0.293 \\
  \text{2010-12-17} &   0.299 &   0.447 &   0.347 &  -0.104 &  -0.409 &   0.264 \\
  \text{2010-12-24} &  -0.427 &   1.045 &   1.345 &   1.197 &   1.233 &   0.576 \\
  \text{2010-12-31} &   0.746 &   0.129 &  -0.612 &   0.863 &   2.240 &   0.518 \\
\hline
\end{array}
$
}
\end{table}
\end{minipage}
\end{center}

\newpage
\section{Linear returns}
\label{appendix:linearreturns39}

\vspace*{-2.0ex}
\begin{center}
\begin{minipage}{6.00in}
\begin{table}[H]
\centering
\caption{\label{linr39a}
   [2009-12-31]-denominated weekly returns (\%) for the last 39 weeks of 2010.
}
{ \fontsize{10}{13}\selectfont
  \newcolumntype{.}{D{.}{.}{3}}
$
\begin{array}{|l|.....|.|} \hline
  \multicolumn{1}{|c|}{\rule{0mm}{4mm}\textrm{Friday}}
  & \multicolumn{1}{c}{\texttt{~IEF}} & \multicolumn{1}{c}{\texttt{~IWB}}
  & \multicolumn{1}{c}{\texttt{~IWM}} & \multicolumn{1}{c}{\texttt{~EFA}}
  & \multicolumn{1}{c}{\texttt{~EEM}} & \multicolumn{1}{|c|}{\texttt{~PORTF}} \\
\hline
  \text{2010-04-09} &   0.011 &   1.670 &   2.922 &   0.326 &   1.349 &   0.753 \\
  \text{2010-04-16} &   0.979 &  -0.065 &   1.943 &  -0.778 &  -3.181 &   0.122 \\
  \text{2010-04-23} &  -0.398 &   2.292 &   4.286 &  -0.470 &   1.446 &   0.660 \\
  \text{2010-04-30} &   1.298 &  -2.620 &  -3.981 &  -3.709 &  -2.434 &  -1.521 \\
  \text{2010-05-07} &   1.683 &  -7.285 & -10.099 &  -9.406 &  -9.301 &  -4.676 \\
  \text{2010-05-14} &   0.022 &   2.636 &   6.743 &   1.411 &   3.133 &   1.415 \\
  \text{2010-05-21} &   1.714 &  -4.486 &  -7.209 &  -2.479 &  -5.181 &  -1.814 \\
  \text{2010-05-28} &  -0.389 &   0.556 &   1.895 &  -0.542 &   1.831 &  -0.049 \\
  \text{2010-06-04} &   0.848 &  -2.275 &  -4.319 &  -2.605 &  -2.168 &  -1.264 \\
  \text{2010-06-11} &  -0.207 &   2.406 &   2.216 &   3.672 &   3.759 &   1.808 \\
  \text{2010-06-18} &   0.161 &   2.309 &   2.986 &   3.075 &   2.795 &   1.749 \\
  \text{2010-06-25} &   1.042 &  -3.553 &  -3.484 &  -1.638 &  -0.553 &  -1.466 \\
  \text{2010-07-02} &   1.154 &  -5.330 &  -7.406 &  -3.240 &  -4.075 &  -2.538 \\
  \text{2010-07-09} &  -0.631 &   5.217 &   5.026 &   5.467 &   5.385 &   3.233 \\
  \text{2010-07-16} &   1.378 &  -1.251 &  -3.013 &  -0.884 &  -3.202 &  -0.239 \\
  \text{2010-07-23} &  -0.437 &   3.539 &   6.299 &   3.700 &   6.064 &   2.188 \\
  \text{2010-07-30} &   0.930 &  -0.099 &   0.064 &   0.589 &   0.607 &   0.433 \\
  \text{2010-08-06} &   0.971 &   1.827 &   0.193 &   3.387 &   1.649 &   1.917 \\
  \text{2010-08-13} &   1.370 &  -3.835 &  -6.556 &  -5.596 &  -3.396 &  -2.453 \\
  \text{2010-08-20} &   0.265 &  -0.493 &   0.129 &  -0.736 &   0.825 &  -0.288 \\
  \text{2010-08-27} &  -0.265 &  -0.626 &   0.806 &   0.386 &  -1.286 &  -0.247 \\
  \text{2010-09-03} &  -0.572 &   3.752 &   4.317 &   3.626 &   3.736 &   2.207 \\
  \text{2010-09-10} &  -0.635 &   0.428 &  -0.983 &   0.589 &   0.558 &   0.096 \\
  \text{2010-09-17} &   0.566 &   1.531 &   2.400 &   1.510 &   1.843 &   1.188 \\
  \text{2010-09-24} &   1.166 &   2.105 &   3.139 &   3.221 &   2.668 &   2.055 \\
  \text{2010-10-01} &   0.904 &   0.033 &   1.405 &   0.165 &   3.178 &   0.371 \\
  \text{2010-10-08} &   1.331 &   1.720 &   2.293 &   2.577 &   1.940 &   1.798 \\
  \text{2010-10-15} &  -1.377 &   1.107 &   1.631 &   1.215 &   1.189 &   0.265 \\
  \text{2010-10-22} &   0.266 &   0.579 &   0.049 &  -0.442 &  -1.674 &   0.214 \\
  \text{2010-10-29} &  -0.289 &   0.198 &  -0.033 &  -0.515 &   0.218 &  -0.151 \\
  \text{2010-11-05} &   1.002 &   3.852 &   5.604 &   4.031 &   5.749 &   2.899 \\
  \text{2010-11-12} &  -1.588 &  -2.215 &  -2.810 &  -3.240 &  -5.045 &  -2.252 \\
  \text{2010-11-19} &  -0.986 &   0.132 &   0.678 &   0.608 &   0.242 &  -0.140 \\
  \text{2010-11-26} &   0.104 &  -1.008 &   1.244 &  -4.418 &  -4.147 &  -1.471 \\
  \text{2010-12-03} &  -1.315 &   3.604 &   3.956 &   3.755 &   5.675 &   1.920 \\
  \text{2010-12-10} &  -2.802 &   1.389 &   3.359 &   0.405 &  -1.334 &  -0.324 \\
  \text{2010-12-17} &   0.325 &   0.512 &   0.436 &  -0.110 &  -0.461 &   0.291 \\
  \text{2010-12-24} &  -0.465 &   1.205 &   1.706 &   1.276 &   1.396 &   0.638 \\
  \text{2010-12-31} &   0.813 &   0.150 &  -0.779 &   0.929 &   2.581 &   0.577 \\
\hline
\end{array}
$
}
\end{table}
\end{minipage}
\end{center}

\newpage
\begin{center}
\begin{minipage}{6.00in}
\begin{table}[H]
\centering
\caption{\label{linr39b}
   $\alpha$-denominated weekly returns (\%) for the last 39 weeks of 2010.
}
{ \fontsize{10}{13}\selectfont
  \newcolumntype{.}{D{.}{.}{3}}
$
\begin{array}{|l|.....|.|} \hline
  \multicolumn{1}{|c|}{\rule{0mm}{4mm}\textrm{Friday}}
  & \multicolumn{1}{c}{\texttt{~IEF}} & \multicolumn{1}{c}{\texttt{~IWB}}
  & \multicolumn{1}{c}{\texttt{~IWM}} & \multicolumn{1}{c}{\texttt{~EFA}}
  & \multicolumn{1}{c}{\texttt{~EEM}} & \multicolumn{1}{|c|}{\texttt{~PORTF}} \\
\hline
  \text{2010-04-09} &   0.010 &   1.501 &   2.448 &   0.308 &   1.192 &   0.683 \\
  \text{2010-04-16} &   0.871 &  -0.058 &   1.628 &  -0.734 &  -2.811 &   0.111 \\
  \text{2010-04-23} &  -0.354 &   2.060 &   3.591 &  -0.444 &   1.278 &   0.598 \\
  \text{2010-04-30} &   1.155 &  -2.355 &  -3.336 &  -3.502 &  -2.151 &  -1.379 \\
  \text{2010-05-07} &   1.498 &  -6.548 &  -8.462 &  -8.880 &  -8.220 &  -4.240 \\
  \text{2010-05-14} &   0.020 &   2.369 &   5.650 &   1.332 &   2.769 &   1.283 \\
  \text{2010-05-21} &   1.526 &  -4.032 &  -6.040 &  -2.340 &  -4.579 &  -1.645 \\
  \text{2010-05-28} &  -0.346 &   0.500 &   1.588 &  -0.512 &   1.618 &  -0.045 \\
  \text{2010-06-04} &   0.755 &  -2.045 &  -3.619 &  -2.459 &  -1.916 &  -1.146 \\
  \text{2010-06-11} &  -0.184 &   2.163 &   1.857 &   3.467 &   3.322 &   1.639 \\
  \text{2010-06-18} &   0.143 &   2.075 &   2.502 &   2.903 &   2.470 &   1.585 \\
  \text{2010-06-25} &   0.928 &  -3.194 &  -2.919 &  -1.546 &  -0.489 &  -1.329 \\
  \text{2010-07-02} &   1.027 &  -4.791 &  -6.205 &  -3.059 &  -3.601 &  -2.301 \\
  \text{2010-07-09} &  -0.562 &   4.689 &   4.211 &   5.161 &   4.759 &   2.931 \\
  \text{2010-07-16} &   1.227 &  -1.124 &  -2.525 &  -0.835 &  -2.830 &  -0.217 \\
  \text{2010-07-23} &  -0.389 &   3.181 &   5.278 &   3.493 &   5.359 &   1.983 \\
  \text{2010-07-30} &   0.828 &  -0.089 &   0.054 &   0.556 &   0.536 &   0.393 \\
  \text{2010-08-06} &   0.864 &   1.642 &   0.162 &   3.198 &   1.457 &   1.738 \\
  \text{2010-08-13} &   1.220 &  -3.447 &  -5.493 &  -5.283 &  -3.001 &  -2.224 \\
  \text{2010-08-20} &   0.236 &  -0.443 &   0.108 &  -0.695 &   0.729 &  -0.262 \\
  \text{2010-08-27} &  -0.236 &  -0.563 &   0.675 &   0.364 &  -1.137 &  -0.224 \\
  \text{2010-09-03} &  -0.509 &   3.372 &   3.617 &   3.423 &   3.302 &   2.001 \\
  \text{2010-09-10} &  -0.565 &   0.385 &  -0.824 &   0.556 &   0.493 &   0.087 \\
  \text{2010-09-17} &   0.504 &   1.376 &   2.011 &   1.426 &   1.629 &   1.077 \\
  \text{2010-09-24} &   1.038 &   1.892 &   2.630 &   3.041 &   2.358 &   1.863 \\
  \text{2010-10-01} &   0.805 &   0.030 &   1.177 &   0.156 &   2.809 &   0.336 \\
  \text{2010-10-08} &   1.185 &   1.546 &   1.921 &   2.433 &   1.714 &   1.630 \\
  \text{2010-10-15} &  -1.226 &   0.995 &   1.367 &   1.147 &   1.051 &   0.240 \\
  \text{2010-10-22} &   0.237 &   0.520 &   0.041 &  -0.417 &  -1.479 &   0.194 \\
  \text{2010-10-29} &  -0.257 &   0.178 &  -0.028 &  -0.486 &   0.193 &  -0.137 \\
  \text{2010-11-05} &   0.892 &   3.462 &   4.696 &   3.806 &   5.081 &   2.629 \\
  \text{2010-11-12} &  -1.414 &  -1.991 &  -2.354 &  -3.059 &  -4.459 &  -2.042 \\
  \text{2010-11-19} &  -0.878 &   0.119 &   0.568 &   0.574 &   0.214 &  -0.127 \\
  \text{2010-11-26} &   0.093 &  -0.906 &   1.042 &  -4.171 &  -3.665 &  -1.334 \\
  \text{2010-12-03} &  -1.171 &   3.239 &   3.315 &   3.545 &   5.015 &   1.741 \\
  \text{2010-12-10} &  -2.494 &   1.249 &   2.814 &   0.382 &  -1.179 &  -0.294 \\
  \text{2010-12-17} &   0.289 &   0.460 &   0.365 &  -0.104 &  -0.407 &   0.264 \\
  \text{2010-12-24} &  -0.414 &   1.083 &   1.429 &   1.205 &   1.234 &   0.579 \\
  \text{2010-12-31} &   0.724 &   0.135 &  -0.653 &   0.877 &   2.281 &   0.523 \\
\hline
\end{array}
$
}
\end{table}
\end{minipage}
\end{center}


\begin{thebibliography}{2}
\providecommand{\natexlab}[1]{#1}
\providecommand{\url}[1]{\texttt{#1}}
\expandafter\ifx\csname urlstyle\endcsname\relax
  \providecommand{\doi}[1]{doi: #1}\else
  \providecommand{\doi}{doi: \begingroup \urlstyle{rm}\Url}\fi

\bibitem[Markowitz(1987)]{Markowitz:1987wd}
Harry~M. Markowitz.
\newblock \emph{Mean-Variance Analysis in Portfolio Choice and Capital
  Markets}.
\newblock Blackwell, 1987.

\bibitem[Norton(2010)]{Norton:2010fk}
Vic Norton.
\newblock Adjusted closing prices, 2010.
\newblock
  $<$\url{http://vic.norton.name/finance-math/adjustedClosingPrices.pdf}$>$.
\newblock Unpublished manuscript.

\end{thebibliography}
\end{document}